\documentclass[aps,prd,notitlepage,showpacs,nofootinbib,superscriptaddress,twocolumn,nofloats]{revtex4-1}

\usepackage{graphicx}
\usepackage[utf8]{inputenc} 
\usepackage{amsmath}
\usepackage{amssymb}
\usepackage{float}
\usepackage{comment}
\usepackage{slashed}
\usepackage[normalem]{ulem}

\usepackage[T1]{fontenc} 
\usepackage{parskip}
\usepackage{bm}         
\usepackage{array}
\usepackage{multirow}   
\usepackage{booktabs}   

\usepackage{textcomp}
\usepackage{amssymb}

\usepackage{enumerate}
\usepackage{caption}
\usepackage{subcaption}
\usepackage{ragged2e}
 \DeclareCaptionJustification{justified}{\justifying}
\usepackage{amsmath}
\usepackage{amssymb}

\usepackage[usenames,dvipsnames]{xcolor}
\usepackage[colorinlistoftodos]{todonotes}
\usepackage[colorlinks=true,citecolor=darkred,urlcolor=darkred, pdfborder={0 0 0}]{hyperref}
\usepackage[normalem]{ulem}

\definecolor{darkred}{rgb}{0.6,0,0}
\usepackage[colorinlistoftodos]{todonotes}

\definecolor{linkcolor}{rgb}{0,0,0.5}
 
\newcommand {\ignore}[1]{}

\definecolor{bostonuniversityred}{rgb}{0.8, 0.0, 0.0}

\newcommand{\ro}[1]{{\mathrm{#1}}}

\def\gsim{\raise0.3ex\hbox{$\;>$\kern-0.75em\raise-1.1ex\hbox{$\sim\;$}}}
\def\lsim{\raise0.3ex\hbox{$\;<$\kern-0.75em\raise-1.1ex\hbox{$\sim\;$}}}
\def\lfv{lepton flavour violation }

\def\SM{$\mathrm{SU(3)_c \otimes SU(2)_L \otimes U(1)_Y}$ }

\usepackage{soul}

\definecolor{mightnightblue}{RGB}{25,25,112}

\definecolor{brown}{rgb}{0.59, 0.29, 0.0}

\def\lfv{lepton flavour violation }

\def\vev#1{\left\langle #1\right\rangle}
\def\SM{$\mathrm{SU(3)_c \otimes SU(2)_L \otimes U(1)_Y}$ }
\def\21{$\mathrm{SU(2)_L \otimes U(1)_Y}$}
\def\lfv{lepton flavor violation }

\newcommand{\AddrAHEP}{%
  AHEP Group, Institut de F\'{i}sica Corpuscular --
  C.S.I.C./Universitat de Val\`{e}ncia, Parc Cient\'ific de Paterna.\\
 C/ Catedr\'atico Jos\'e Beltr\'an, 2 E-46980 Paterna (Valencia) - SPAIN}

\begin{document}

\title{\boldmath \color{BrickRed} Gravitational footprints of massive neutrinos and lepton number breaking}

\author{Andrea Addazi}
\email{andrea.addazi@lngs.infn.it}
\affiliation{Department of Physics \& Center for Field Theory and Particle Physics, Fudan University, 
200433 Shanghai, China}
\affiliation{College of Physics, Sichuan University, Chengdu, 610065, China}

\author{Antonino Marcian\`o}
\email{marciano@fudan.edu.cn}
\affiliation{Department of Physics \& Center for Field Theory and Particle Physics, Fudan University, 
200433 Shanghai, China}

\author{Ant{\'o}nio~P.~Morais}
\email{aapmorais@ua.pt}
\affiliation{Departamento de F\'isica, Universidade de Aveiro and CIDMA, 
Campus de Santiago, 3810-183 Aveiro, Portugal, EU}

\author{Roman~Pasechnik}
\email{Roman.Pasechnik@thep.lu.se}
\affiliation{Department of Astronomy and Theoretical Physics, Lund University, 
221 00 Lund, Sweden, EU}

\author{Rahul Srivastava}
\email{rahulsri@ific.uv.es}
\affiliation{\AddrAHEP}

\author{Jos\'{e} W. F. Valle}
\email{valle@ific.uv.es}
\affiliation{\AddrAHEP}

\begin{abstract}
\vspace{0.5cm}
\noindent
We investigate the production of primordial Gravitational Waves (GWs) arising from First Order Phase Transitions (FOPTs) 
associated to neutrino mass generation in the context of type-I and inverse seesaw schemes. We examine both ``high-scale'' 
as well as ``low-scale'' variants, with either explicit or spontaneously broken lepton number symmetry $U(1)_L$ in 
the neutrino sector. In the latter case, a pseudo-Goldstone majoron-like boson may provide a candidate for cosmological 
dark matter. We find that schemes with softly-broken $U(1)_L$ and with single Higgs-doublet scalar sector lead 
to either no FOPTs or too weak FOPTs, precluding the detectability of GWs in present or near future measurements. 
Nevertheless, we found that, in the majoron-like seesaw scheme with spontaneously broken $U(1)_L$ at finite temperatures, 
one can have strong FOPTs and non-trivial primordial GW spectra which can fall well within the frequency and amplitude 
sensitivity of upcoming experiments, including LISA, BBO and u-DECIGO. However, GWs observability clashes 
with invisible Higgs decay constraints from the LHC. A simple and consistent fix is to assume the majoron-like mass to lie above
the Higgs-decay kinematical threshold. We also found that the majoron-like variant of the low-scale seesaw mechanism 
implies a different GW spectrum than the one expected in the high-scale seesaw. This feature will be testable 
in future experiments. Our analysis shows that GWs can provide a new and complementary portal to test 
the neutrino mass generation mechanism.
\end{abstract}

\maketitle
\noindent

\section{Introduction}
\label{Sec:intro}

Non-zero neutrino masses constitute one of the most robust evidences for new
physics~\cite{Kajita:2016cak,McDonald:2016ixn,Valle:2015pba}. 
Despite great efforts over the last two decades to underpin the origin of neutrino mass, the basic underlying 
mechanism remains as elusive as ever. Small neutrino masses can be generated in many ways, both for
Majorana~\cite{weinberg:1980bf,Boucenna:2014zba} and
Dirac~\cite{Ma:2014qra,Ma:2015mjd,Chulia:2016ngi,CentellesChulia:2018gwr,CentellesChulia:2018bkz, Bonilla:2018ynb} 
neutrinos. Here, we focus on the various variants of the popular type-I seesaw mechanism 
for Majorana neutrinos~\cite{Minkowski:1977sc,Yanagida:1979as,Mohapatra:1979ia,Schechter:1980gr}. We consider 
both high- and low-scale~\cite{Mohapatra:1986bd,GonzalezGarcia:1988rw,Akhmedov:1995ip,Akhmedov:1995vm,Malinsky:2005bi} 
realizations, with explicit or spontaneous lepton number violation, in which \SM singlet neutrinos act as neutrino 
mass mediators. Besides oscillation and neutrinoless double beta decay ($0 \nu 2 \beta$) searches, neutrino masses 
can be probed through Charged Lepton Flavor Violation (CLFV) experiments 
at the high intensity and/or high energy frontier~\cite{Das:2012ii,Deppisch:2013cya,Deppisch:2015qwa}. Moreover, 
neutrino mass generation can leave signatures at high-energy colliders like the Large Hadron Collider
(LHC)~\cite{Joshipura:1992hp,Diaz:1998zg,Bonilla:2015jdf,Bonilla:2015uwa}. 

The detection of Gravitational Waves (GWs) by the LIGO team has opened an entirely novel method to probe the underlying new physics 
associated to neutrino mass generation. It was advocated that the spectrum of primordial GWs, potentially measurable at the currently 
planned GW interferometers, may represent an important cutting-edge probe for new physics. This follows from the fact that these 
interferometers can be sensitive enough to measure the echoes of the possible First Order Phase Transitions (FOPTs), which might 
have happened in the past cosmological history~\cite{Kosowsky:1992rz}. 

In this letter, we focus on possible gravitational footprints of the various variants of the popular type-I and inverse 
seesaw mechanisms for Majorana neutrinos. The relevant part of the minimal type-I seesaw Lagrangian is given by
\begin{eqnarray}
   \mathcal{L}_{\rm Yuk}^{\rm Type-I} & = & Y_\nu \bar{L} H \nu^c + M \nu^c \nu^c + h.c.
 \label{lag-typeI}
\end{eqnarray}
Here, $L = (\nu, l)^T$ are the SM lepton doublets, $H$ is the SM Higgs doublet, $\nu^c$ are the three SM singlet ``right-handed'' 
neutrinos. The $3 \times 3$ matrices $Y_\nu$ and $M$ are the Yukawa coupling and the $\nu^c$ mass matrix, respectively. 
Due to the Pauli principle the latter is symmetric. Notice that, for brevity, we omit family indices throughout this letter. 
Notice also that the mass term explicitly breaks the lepton number symmetry $U(1)_L$ to its $\mathbb{Z}_2$ subgroup. 
The electroweak (EW) symmetry is broken by the vacuum expectation value (vev) of the Higgs field, i.e. 
$\vev{H} = v_h / \sqrt{2}$,  generating the light neutrino masses
\begin{eqnarray}
 m_\nu^{\rm Type-I} & = & \frac{v_h^2}{2} Y_\nu^T M^{-1} Y_\nu .
 \label{mass-typeI}
\end{eqnarray}
The lightness of the left-handed neutrinos is then ascribed to the heaviness of the ``right-handed'' isosinglet partners 
e.g. for $Y \sim \mathcal{O}(1), \, M \sim \mathcal{O}(10^{14})$ GeV, one gets $m_\nu \sim \mathcal{O} (0.1)$ eV.

Another popular realization of this idea is the ``low-scale'' variant, in which two gauge singlet fermions $\nu^c$ and $S$ 
are added sequentially to the SM particle
content~\cite{Mohapatra:1986bd,GonzalezGarcia:1988rw,Akhmedov:1995ip,Akhmedov:1995vm,Malinsky:2005bi}.  
The template of these schemes has exact conservation of lepton number and, as a result, strictly massless neutrinos. Yet flavor 
is violated to a potentially large degree, subjected only to constraints from weak interaction precision observables, such 
as universality tests~\cite{Bernabeu:1987gr,Branco:1989bn,Rius:1989gk,Deppisch:2004fa,Antusch:2006vw}. To this template 
one adds a small seed of lepton number violation, leading to nonzero neutrino mass. One example is the so-called 
``inverse seesaw'' mechanism, where the smallness of the neutrino mass is linked to the breaking of the lepton 
number symmetry $U(1)_L$ to its $\mathbb{Z}_2$ subgroup, through the so called $\mu$-term. The relevant part 
of Lagrangian in this case is given by
\begin{equation}
\label{lag-inv}
\mathcal{L}_{\rm Yuk}^{\rm Inverse} =  Y_\nu \bar{L} H \nu^c \, + \, M \nu^c S \, + \, \mu S S \, + \, \rm{h.c.} \, ,
\end{equation}
where $\mu$ is also a $3 \times 3$ symmetric matrix. The light neutrino mass is then given by 
\begin{eqnarray}\label{mass-inv}
m_\nu^{\rm Inverse} = \frac{v_h^2}{2} Y_\nu^T M^{T^{-1}} \mu M^{-1} Y_\nu ~.
\end{eqnarray}
Note that small neutrino masses are ``protected'', since $m_\nu\to 0$ as the lepton number symmetry gets restored by having 
$\mu\to 0$~\cite{Mohapatra:1986bd,GonzalezGarcia:1988rw,Akhmedov:1995ip,Akhmedov:1995vm,Malinsky:2005bi}.
In this case there can be sizable unitarity violation in neutrino propagation~\cite{Escrihuela:2015wra,Forero:2011pc,Miranda:2016wdr}. 

For both high- and low-scale seesaw, one can have spontaneous breaking of $U(1)_L \to \mathbb{Z}_2$, leading to the so-called 
majoron variants of the seesaw~\cite{Chikashige:1980ui,Schechter:1981cv,GonzalezGarcia:1988rw}. This is accomplished by adding 
the SM singlet scalar $\sigma$, which carries two units of lepton number charge. Then $\vev{\sigma} \equiv v_\sigma$ spontaneously 
breaks $U(1)_L \to \mathbb{Z}_2$, leading to a dynamical 
explanation of the small neutrino masses. To get the majoron variants of minimal type-I and inverse seesaw 
one should replace
\begin{eqnarray}
M \to Y_\sigma  \, v_\sigma/\sqrt{2}\,, \qquad \mu \to Y_\sigma \, v_\sigma/\sqrt{2}
\label{eq:majoron}
\end{eqnarray}
in Eq.~\eqref{lag-typeI} and \eqref{lag-inv}, respectively. An additional attractive feature of majoron models is the existence 
of a pseudo Nambu-Goldstone boson commonly dubbed as majoron. The latter carries odd charge under $\mathbb{Z}_2$ thus
providing a good~\cite{Berezinsky:1993fm,Lattanzi:2007ux,Kuo:2018fgw}, and testable~\cite{Lattanzi:2013uza,Bazzocchi:2008fh} 
dark matter candidate. In the standard majoron seesaw schemes, the majoron mass is considered to be small, of order keV, 
for it to be a suitable warm dark matter candidate. However, the current stringent constraints on invisible Higgs boson decay 
modes \cite{Bonilla:2015jdf,Bonilla:2015uwa,Aaboud:2018sfi,Sirunyan:2018owy} put severe limitation 
on how large $h\to \sigma \sigma$ coupling could be. On the other hand, the strength of the cosmological phase transition 
is expected to be strongly correlated with the size of the Higgs-majoron quartic coupling. It remains an open question whether 
it is possible to reconcile the current LHC bounds on the invisible Higgs decays in the inverse seesaw scenario 
featuring a keV-scale majoron dark matter with the existence of strong EW FOPTs yielding potentially observable GWs signals.

In this work, we consider the case where the lepton number symmetry is broken also explicitly, but softly. This way the majoron can pick up a mass.
  If this is larger than a half of the Standard Model (SM) Higgs boson mass, $m_h/2 \simeq 62.5~\mathrm{GeV}$, 
  the invisible Higgs decays will be kinematically forbidden, avoiding the stringent constraints on the Higgs-majoron quartic coupling. In such scenario
  the heavy majoron-like state can provide a viable candidate for cold dark matter (CDM). The same type of scalar CDM scenarios 
with exact $\mathbb{Z}_2$ parity have been broadly studied in singlet extensions of the SM \cite{Costa:2014qga,Burgess:2000yq,Gonderinger:2012rd,Cline:2013gha,Gabrielli:2013hma,Khan:2014kba,Ghorbani:2019itr,Cosme:2018wfh,Azevedo:2018oxv,Grzadkowski:2018nbc,Chiang:2017nmu}, which share very similar properties with the scalar sector of the model under consideration.

\section{Scalar sector}
\label{Sec:scalar}

The scalar sector of the majoron inverse seesaw model has been extensively studied in the literature, including 
the perturbative unitarity and stability of the scalar potential, as well as the electroweak precision tests 
and the bounds on Higgs-majoron couplings~\cite{Bonilla:2015uwa,Brune:2018sab}. The scalar potential is written as follows
\begin{eqnarray}
\mathcal{V}_0(\Phi,\sigma) & = & \mu_\Phi^2 \Phi^\dag \Phi + \lambda_\Phi (\Phi^\dag \Phi)^2 
+ \mu_\sigma^2 \sigma^\dag \sigma + \lambda_\sigma (\sigma^\dag \sigma)^2 \nonumber \\
& & + \lambda_{\Phi \sigma} \Phi^\dag \Phi \sigma^\dag \sigma + 
\Big(\frac12 \mu_b^2 \sigma^2 + {\rm h.c.}\Big) \, ,
\label{V0-tot}
\end{eqnarray}
with $\Phi$ and $\sigma$ given by
\begin{align}
\begin{aligned}
\Phi &= \frac{1}{\sqrt{2}} 
\begin{pmatrix} 
G + i G' \\ 
\phi_h + h + i \eta 
\end{pmatrix}\,,
\end{aligned} 
\quad
\begin{aligned}
\sigma = \dfrac{1}{\sqrt{2}}( \phi_\sigma + \sigma_R + i \sigma_I )\,,
\end{aligned} 
\end{align}
where $h$, $\eta$, $G$, $G'$, $\sigma_R$, $\sigma_I$ are real scalars. These latter fields represent quantum fluctuations 
about the classical mean-fields $\phi_{\alpha}=\{\phi_h,\,\phi_\sigma\}$. In the zero-temperature limit, the mean-fields 
approach the corresponding vevs, i.e. $\phi_{h,\sigma}(T=0)\equiv v_{h,\sigma}$, where $v_h=246$ GeV is the SM Higgs vev.
Besides, one of the physical CP-even scalar states, with a small or no mixing with $\sigma_R$, is identified with 
the SM-like Higgs boson with mass, $m_{h_1}\equiv m_h=125$ GeV. The last (soft) term appearing in Eq.~\eqref{V0-tot} 
implements the explicit breakdown $U(1)_L \to \mathbb{Z}_2$, and hence provides a pseudo-Goldstone mass to the imaginary 
part of the field $\sigma_I$ known as majoron~ \footnote{One could also add other other explicit breaking terms such as
 $\sigma \Phi^\dag \Phi$ but here we stick to the simplest possibility of mass terms.}. 

In what follows, we discuss further implications of the majoron seesaw scenario in both versions of the scalar sector, with explicit (vanishing $v_\sigma$)
and with spontaneous\footnote{More precisely, here we refer to a mixture of explicit and spontaneous $U(1)_L$ breakings in the
scalar sector of the model since
the soft scalar mass term $\mu_b\not=0$ is always present in the scalar potential (\ref{V0-tot}) to provide a non-zero mass 
to the physical majoron state. This should not be confused with the neutrino sector where $U(1)_L$ is {\it always} spontaneously broken by a small or large $v_\sigma$ vev in the considered majoron versions of the seesaw schemes.} ($v_\sigma\not=0$) lepton-number $U(1)_L \to \mathbb{Z}_2$ symmetry breaking, for physics 
of cosmological EW FOPTs and examine the associated GWs spectra. In the first version, no mixing occurs so that
\begin{eqnarray}
&& m_{h_1}^2 = 2\lambda_h v_h^2 \,, \qquad m_{h_2}^2 = \mu_\sigma^2 + \mu_b^2 + \frac{\lambda_{\sigma h} v_h^2}{2}\,, \\
&& m_A^2 = \mu_\sigma^2 - \mu_b^2 + \frac{\lambda_{\sigma h} v_h^2}{2} \,,
\label{mA}
\end{eqnarray}
for the SM Higgs boson, CP-even and CP-odd (majoron) scalars, respectively. In the second version, the physical CP-even 
states acquire masses, 
\begin{eqnarray}
m_{h_1,h_2}^2=\lambda_h v_h^2 + \lambda_\sigma v_\sigma^2 \mp 
\frac{\lambda_\sigma v_\sigma^2 - \lambda_h v_h^2}{\cos 2\theta} \,,
\end{eqnarray}
in terms of $h$-$\sigma_R$ mixing angle $\theta$, while the majoron gets a pseudo-Goldstone mass,
\begin{eqnarray}
m_A^2\equiv m_{\sigma_I}^2 = - 2\mu_b^2 \,, \qquad \mu_b^2 < 0 \,.
\end{eqnarray}
While in the latter case, a very light majoron (compared to the EW scale $v_h$) would imply 
setting an equally small $\mu_b$ parameter, in the former case, Eq.~\eqref{mA}, this limit 
relies on a very 
strong fine tuning between $\mu_\sigma$, $\mu_b$ and $\lambda_{\sigma h}$ model parameters. 
In the numerical analysis of GW spectra below we do not enforce the majoron mass $m_A$ 
to be small (e.g.~at a keV scale) and treat it as a free parameter instead.

\section{Gravitational waves from FOPTs}
\label{Sec:GWs}

In order to characterize the features of the GWs originating from FOPTs in seesaw schemes, we calculate the strenght of the phase transition $\alpha$ at the bubble nucleation temperature $T_n$ typically defined through the trace anomaly as \cite{Hindmarsh:2015qta,Hindmarsh:2017gnf}
\begin{equation}
\alpha = \frac{1}{\rho_\gamma} \Big[ V_i - V_f - \dfrac{T_n}{4} \Big( \frac{\partial V_i}{\partial T} - 
\frac{\partial V_f}{\partial T} \Big) \Big] \,,
\label{alpha}
\end{equation}
with
\begin{equation}
\rho_\gamma = g_* \frac{\pi^2}{30} T_n^4
\end{equation}
being the energy density of the radiation medium at the bubble nucleation epoch found in terms of the number 
of relativistic d.o.f.'s.~$g_* \simeq 106.75$~\cite{Grojean:2006bp,Leitao:2015fmj,
Caprini:2015zlo,Caprini:2019egz}. Above, $V_i$ and $V_f$ are the values of the effective potential in the symmetric and broken phases just before and after the transition takes place, respectively. Another key quantity to calculate the GW spectrum is the inverse time-scale $\beta$ of the phase transition, which, in units of the Hubble parameter $H$, reads as 
\begin{equation}
\frac{\beta}{H} = T_n  \left. \frac{\partial}{\partial T} \left( \frac{\hat{S}_3}{T}\right) \right|_{T_n}\,,
\label{betaH}
\end{equation}
where $\hat{S}_3$ is the Euclidean action. In this work, we do not consider the case of runaway bubbles and use the formalism outlined in Ref.~\cite{Caprini:2019egz} to calculate the spectrum of primordial GWs.

For the case of non-runaway nucleated bubbles, the intensity of the GW radiation grows with the ratio $\Delta v_n/T_n$, where
\begin{equation}
    \Delta v_n^\phi = |v_\phi^f - v_\phi^i|\,, \qquad \phi = h,\sigma
\end{equation}
defines the difference between the VEVs of the initial (metastable) and final (stable) phases at the bubble nucleation temperature $T_n$. The quantity $\Delta v_n/T_n$ also offers a measure of the strength of the phase transition alternative to $\alpha$. However, while the latter is of common use in the context of GWs, the former is typically referred to in the context of electroweak baryogenesis. In this work, we will consider both quantities on the same footing.

From the discussion in Ref.~\cite{Hindmarsh:2017gnf,Ellis:2019oqb}, it follows that bubble wall collisions do not provide an efficient way of producing GWs in the models of interest to us here. As a result, GWs originate mainly from two sources:
\begin{enumerate}[I.]
\item Magnetohydrodynamic (MHD) turbulence;
\item Sound shock waves (SW) of the early Universe plasma, generated by the bubble's violent expansion.
\end{enumerate}
These contributions arise over transient times in the early Universe and get subsequently ``redshifted'' by the expansion. 
To a present observer this appears as a cosmic gravitational stochastic background. Intuitively, one expects that from any 
of these leading order contributions, a high wall velocity is necessary to generate detectable GWs. In our numerical analysis,
performed with the help of the \texttt{CosmoTransitions} package \cite{Wainwright:2011kj}, we have considered supersonic detonations such that the bubble wall velocity maximizes the GW peak amplitude and is above the Chapman-Jouguet velocity defined as
\begin{equation}
	v_\mathrm{J} = \dfrac{1}{1+\alpha} \left(c_s + \sqrt{\alpha^2 + \tfrac{2}{3} \alpha}\right)\,,
	\label{eq:vJ}
\end{equation}
with $c_s = \tfrac{1}{\sqrt{3}}$ being the speed of sound. Besides, in our results the SW contribution dominates the peak frequency and the peak amplitude. Furthermore, the state of the art expressions derived in Ref.~\cite{Caprini:2019egz} do not account for MHD-turbulence effects due to large theoretical uncertainties. Therefore, in the remainder of this work, we will not take such effects into consideration. Note, for certain parameter configurations one also expects sequential phase transition patterns leading to multi-peak GWs spectra studied for the first time in Refs.~\cite{Vieu:2018zze,Morais:2019fnm}.

\subsection{Seesaw-induced GWs spectra}
\label{Sec:seesaw-GWs}

To investigate the eventual occurrence of phase transitions, the standard way is to incorporate into the effective potential 
the tree-level zero temperature components, the Coleman-Weinberg corrections, the full one-loop finite-temperature corrections, 
as well as the Daisy re-summation. It is worth noticing that within the type-I seesaw mechanism with explicitly broken 
lepton number, no FOPTs are obtained. The heavy isosinglet neutrinos practically decouple at the EW scale, and do not alter 
the nature of the EW phase transition. In contrast, in the non-majoron inverse seesaw mechanism (i.e. without adding a SM 
singlet scalar) the singlet neutrinos lie closer to the EW scale, and can have a sizable coupling to the Higgs boson. 
However, even by adding a large number of singlet neutrino species it is impossible to generate any sizable FOPTs at 
the quantum level since the fermion thermal loop contributions generate highly suppressed terms to the effective 
potential in the high-temperature expansion. Indeed, we find relatively weak FOPTs for many points in parameter space 
in this case and the corresponding GW ``intensity'' parameter $h^2 \Omega^\ro{peak}_\ro{GW}$ lies far below the sensitivity 
of any conceivable experiment\footnote{We emphasize that variations of the Yukawa coupling $Y_\nu$ in the range $1\div 10$ 
cannot allow the detectability of the GW signal in the type-I seesaw scenario.}.

In the majoron inverse seesaw scenarios, i.e.~when the scalar sector of the SM is extended by incorporating an additional 
complex SM-singlet scalar state, the situation changes dramatically. Indeed, it is well known that the presence of additional 
SM scalar singlets significantly enhances the strength and multiplicity of the FOPTs, 
and in some cases leads 
to potentially detectable GW spectra (see e.g.~Refs.~\cite{Hashino:2018wee,Alves:2018jsw,Kurup:2017dzf,Hashino:2016xoj,Kakizaki:2015wua}). The presence of an additional classical field $\phi_\sigma\not=0$ coupled 
substantially to the Higgs boson strongly affects the shape of the effective potential at non-zero temperature $T$ 
allowing for a richer pattern of EW FOPTs. Note, in the considered majoron seesaw this happens in both variants with 
explicit and spontaneous $U(1)_L$ breaking at $T=0$ discussed above in Sec.~\ref{Sec:scalar}.

The values of the coupling constants used in our numerical analysis satisfy the conservative bounds provided by tree-level 
perturbativity, $|\lambda_i|<4 \pi$ and $|Y_i| < \sqrt{4\pi}$, for the quartic and Yukawa couplings, 
respectively\footnote{For a more involved analysis of the perturbativity constraints in a scalar extension of the 
SM, see e.g. Ref.~\citep{Moretti:2015cwa}}. Since the tree-level potential also receives both quantum and finite 
temperature corrections, we only considered values within the ranges $|\lambda_i|<5$ and $|Y_i| < 3.5$, even more 
conservative bounds in a zero-temperature theory.

\begin{figure}[!b]
\centering
\begin{center}
\includegraphics[height=5cm,width=8cm]{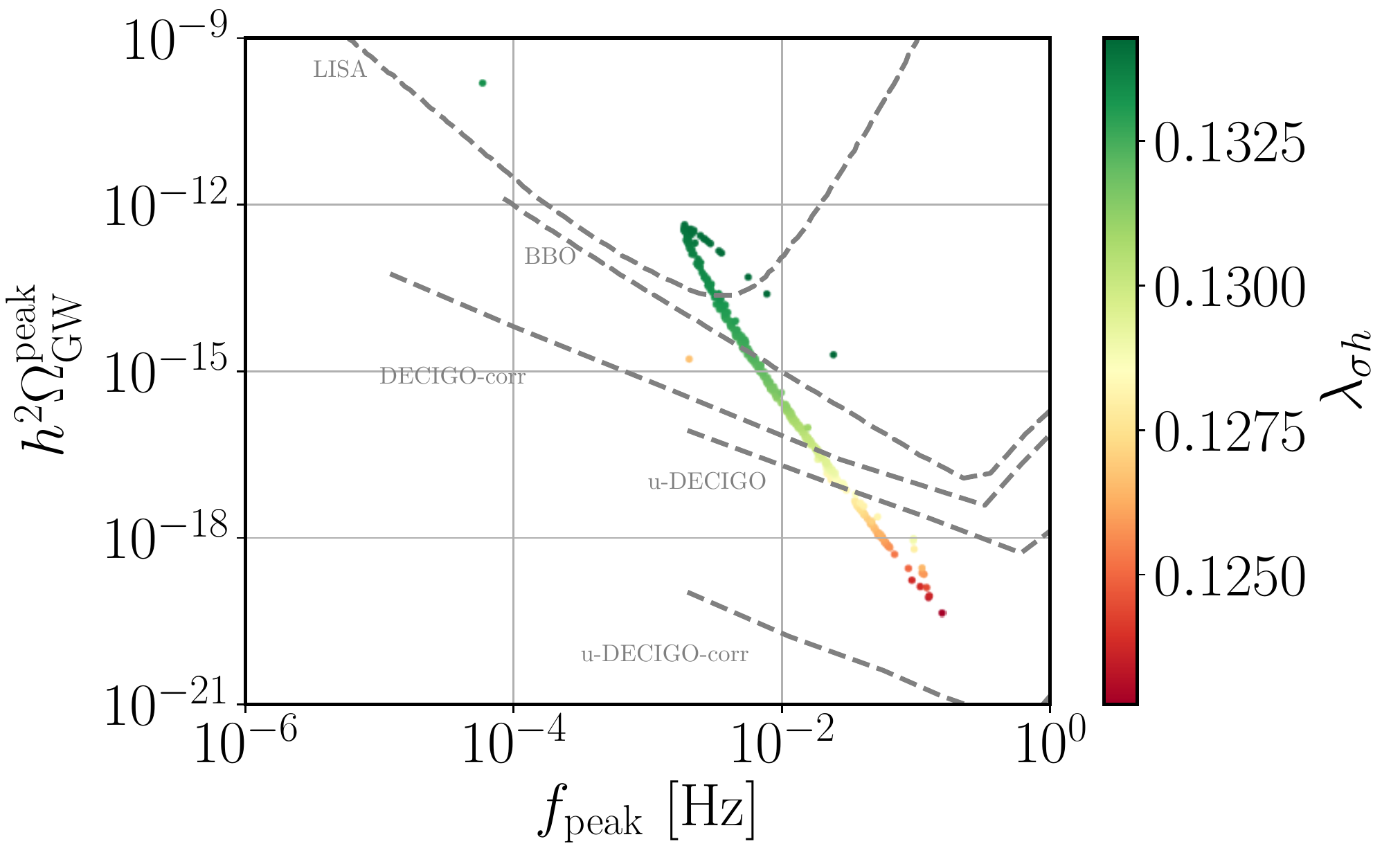}
 \end{center}
\caption{\footnotesize 
The GW spectrum as a function of $\lambda_{\sigma h} $ for the case 
of spontaneously broken $U(1)_L$ symmetry. No solutions consistent with
the LHC bound on invisibly-decaying Higgs were found.
Other model parameters are fixed as: 
$m_A = 1$ keV, $m_{h_2} = 591$ GeV,
$v_\sigma = 858$ GeV, $Y_{\sigma,1}= 1.20$, and $Y_{\sigma,2}= 1.66$.
}
\label{fig:kev-case}
\end{figure}
Due to the current LHC constraints on invisible Higgs decays~\cite{Bonilla:2015jdf,Bonilla:2015uwa,Aaboud:2018sfi,Sirunyan:2018owy}
one has a bound $\lambda_{\sigma h} \lsim 0.03$ the Higgs-majoron quartic coupling in the case of light keV-scale majoron.
  Under this assumption in our numerical scan we did not find any point with an EW FOPT that is strong enough for a potential
  observability of the resulting GW spectra.
This is illustrated in Fig~\ref{fig:kev-case}.

There is a strong correlation of the peak-amplitude with
$\lambda_{\sigma h}$ value such that requiring the latter to be very small makes the GW signals well below the reach 
of LISA or even the planned BBO and DECIGO missions. Note, this is the case for both considered versions of the
majoron inverse seesaw model, with explicit and spontaneous lepton number symmetry breaking in the scalar sector 
at $T=0$. Therefore, we conclude that the standard keV-scale (warm) majoron dark matter scenario associated with 
the inverse seesaw mechanism cannot be probed by the GW astrophysics in the current simplest formulation. 
For this reason, from now on we only consider the case of heavy majoron $m_A>m_h/2$ kinematically closing 
the invisible Higgs decay channel and thus enabling us to consider larger values 
of $\lambda_{\sigma h}$ that ensure the existence of the strong FOPTs in the model under consideration.

\subsection{Inverse seesaw with majoron: small $v_\sigma$ case}
\label{Sec:vs_0}

Let us now consider the case of a genuine inverse seesaw with majoron and very small singlet VEV $v_\sigma$, effectively generating an equally small $\mu S S$ term in the Lagrangian (\ref{lag-inv}). As discussed above, this mechanism offers a dynamical explanation for light neutrino masses, whose scale can be attributed to the (tiny) scale of spontaneous breaking of the lepton-number $\mathrm{U}(1)_L$ symmetry in the neutrino sector (while being softly-broken in the scalar sector).

\begin{figure}[htb]
\centering
\begin{center}
\includegraphics[height=5cm,width=8cm]{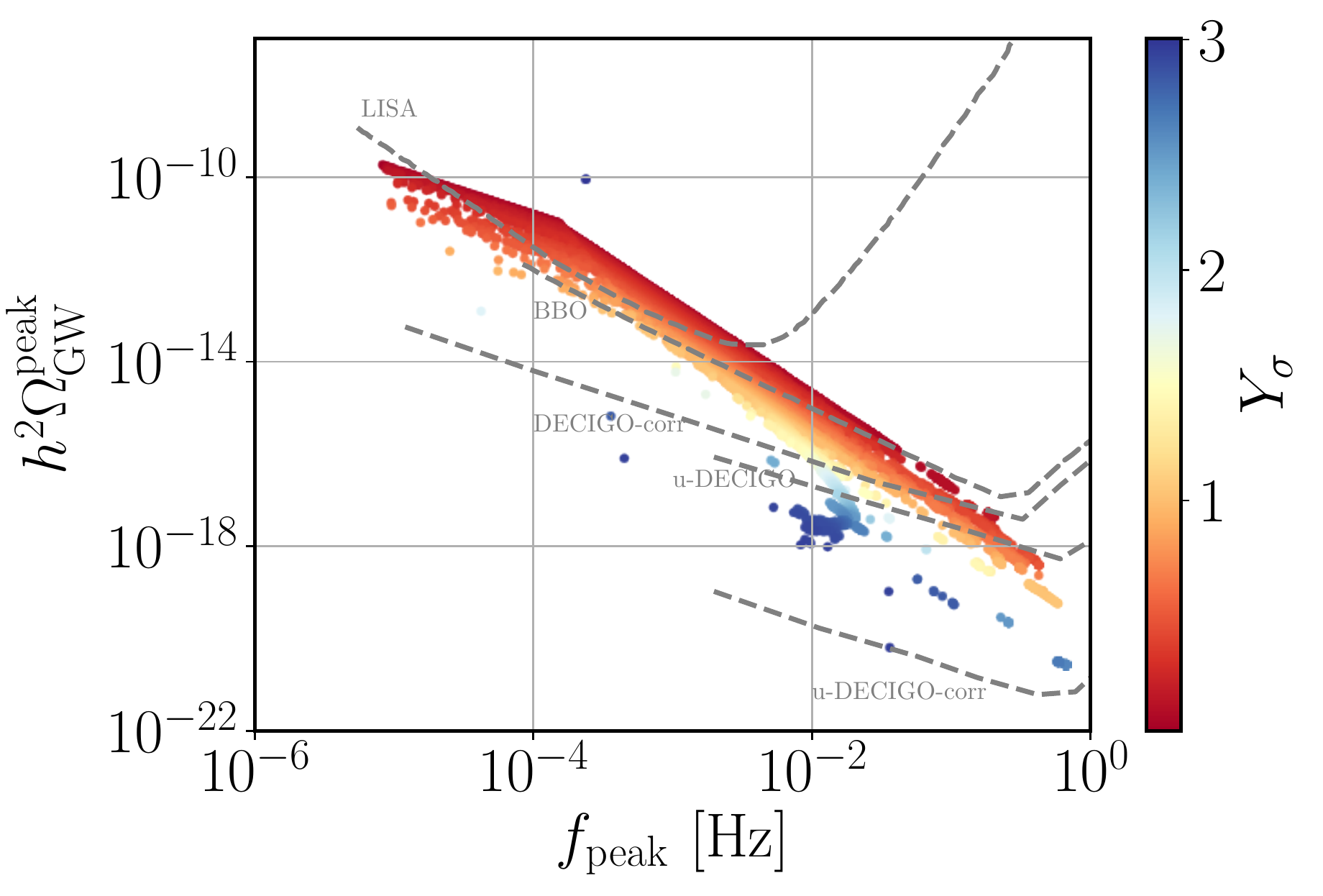}
 \end{center}
\caption{\footnotesize 
The GW spectrum as a function of the Yukawa $Y_\sigma$ coupling in the case 
of softly-broken $U(1)_L$ symmetry (i.e.~$v_\sigma=0$). 
Order one variation of $Y_\sigma$ correspond to several order of magnitude 
variations in the GW power spectrum. Other model parameters are fixed as 
$\lambda_\sigma\!=\!0.37$, 
$\lambda_{\sigma h} \!=\!1.07$, 
$M\!=\!239.4$ GeV, 
$m_{h_2}\!=\!154.6$ GeV and 
$m_A\!=\!369.9$ GeV.}
\label{fig:Ysigma}
\end{figure}

In order to understand the role of heavy neutrino in the generation of the GW spectra in the majoron inverse seesaw
scenario, we study the sensitivity of the GW peak-amplitude $h^2 \Omega^\ro{peak}_\ro{GW}$ originated by the EW FOPTs 
with respect to the variation of the Yukawa coupling $Y_{\sigma}$. As shown in Fig.~\ref{fig:Ysigma} an order one
variation in the Yukawa coupling reflects into violent variations by several orders of magnitude in the GWs spectrum, 
with all the other parameters fixed. The results are shown together with the projected sensitivities expected 
in LISA, and the planned u-DECIGO and BBO missions \citep{Caprini:2019egz,Caprini:2015zlo,Audley:2017drz,Kudoh:2005as}. 
We have taken the u-DECIGO sensitivity curves from Ref.~\citep{Nakayama:2009ce}, whereas the sensitivities 
of other experiments are taken from Ref.~\citep{Moore:2014lga}. Here, for simplicity we have used the soft 
$U(1)_L$ breaking scenario at zero temperature, with vanishing $v_\sigma$ providing a tiny value of $\mu$ 
in the inverse seesaw mechanism according to Eq.~(\ref{eq:majoron}). 
We find a large number of points with strong FOPTs that generate the GW peak-amplitudes well within 
the projected LISA sensitivity, with typical values of $Y_{\sigma}$ below unity.

As is typical in models with several scalars, due to the presence of two classical fields $\{\phi_h,\,\phi_\sigma\}$
in the effective potential at finite temperatures, besides a plenty of single-step FOPTs one also finds double-step and, 
in some rare cases, even triple-step FOPTs for a given parameter space point and at well-separated nucleation temperatures. 
One could naturally expect the presence of several peaks in the corresponding GW spectrum associated with each FOPTs in such
a chain of transitions. Notice also that the analyses of GW spectra for the multi-step FOPT scenarios involving EW phase
transitions require particular care, as discussed for instance in Ref.~\cite{Ellis:2018mja}.
\begin{figure}[htb]
\begin{center}
 \begin{tabular}{lr}
\begin{subfigure}[b]{0.45\textwidth}
\includegraphics[height=5.5cm,width=8.0cm]{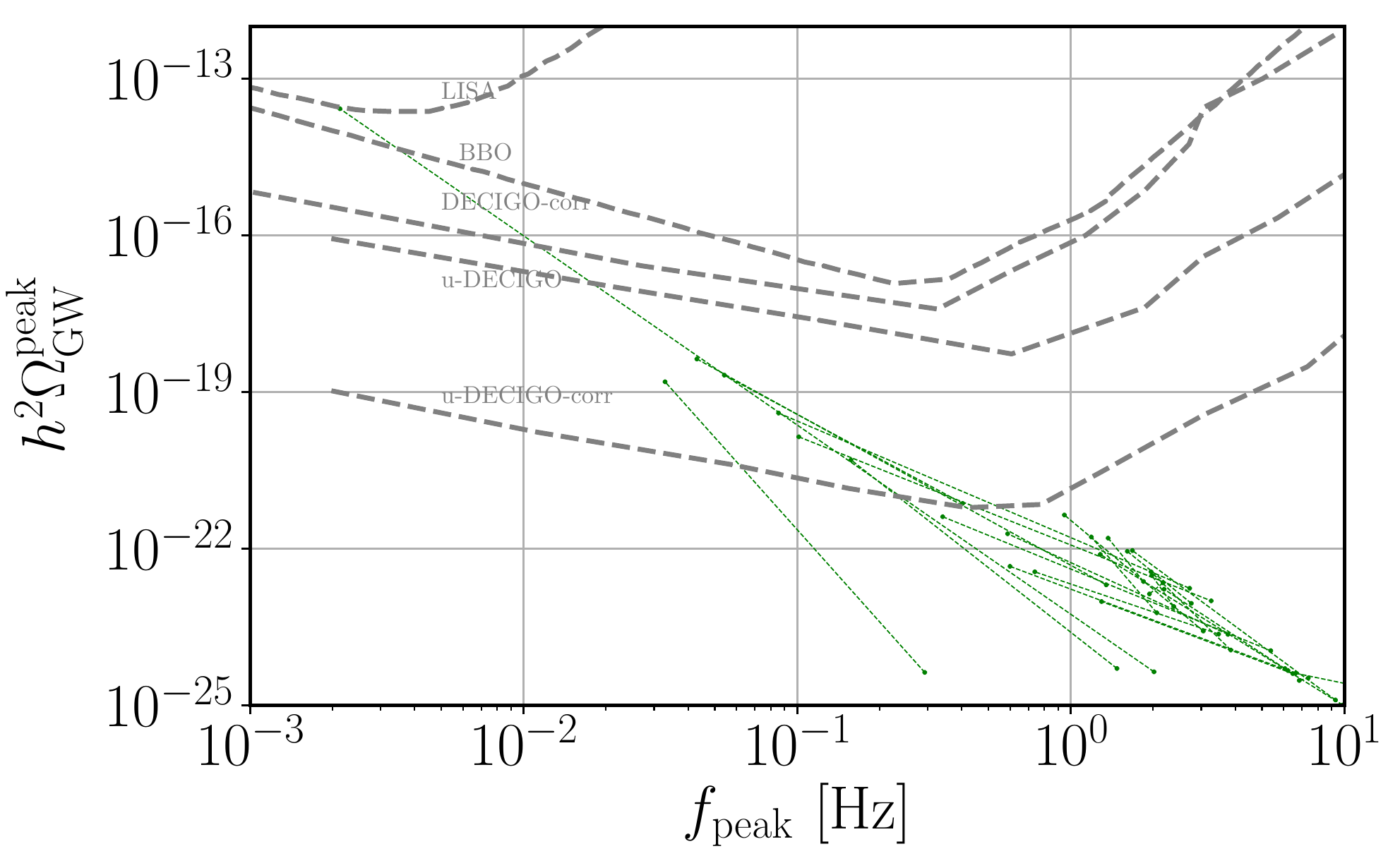}
\caption[a]{\footnotesize  Selected double-peak scenarios within the LISA and BBO sensitivity ranges. 
The two ends of each line represent the location of the peaks of the double-peak GW spectrum. 
The two maxima in each double-peak GW spectra are joined by a straight line, in order to easily 
identify the peaks associated with each other.}
\label{fig:d-peak-line}
\end{subfigure}
\\
\begin{subfigure}[b]{0.45\textwidth}
\includegraphics[height=5.5cm,width=8.0cm]{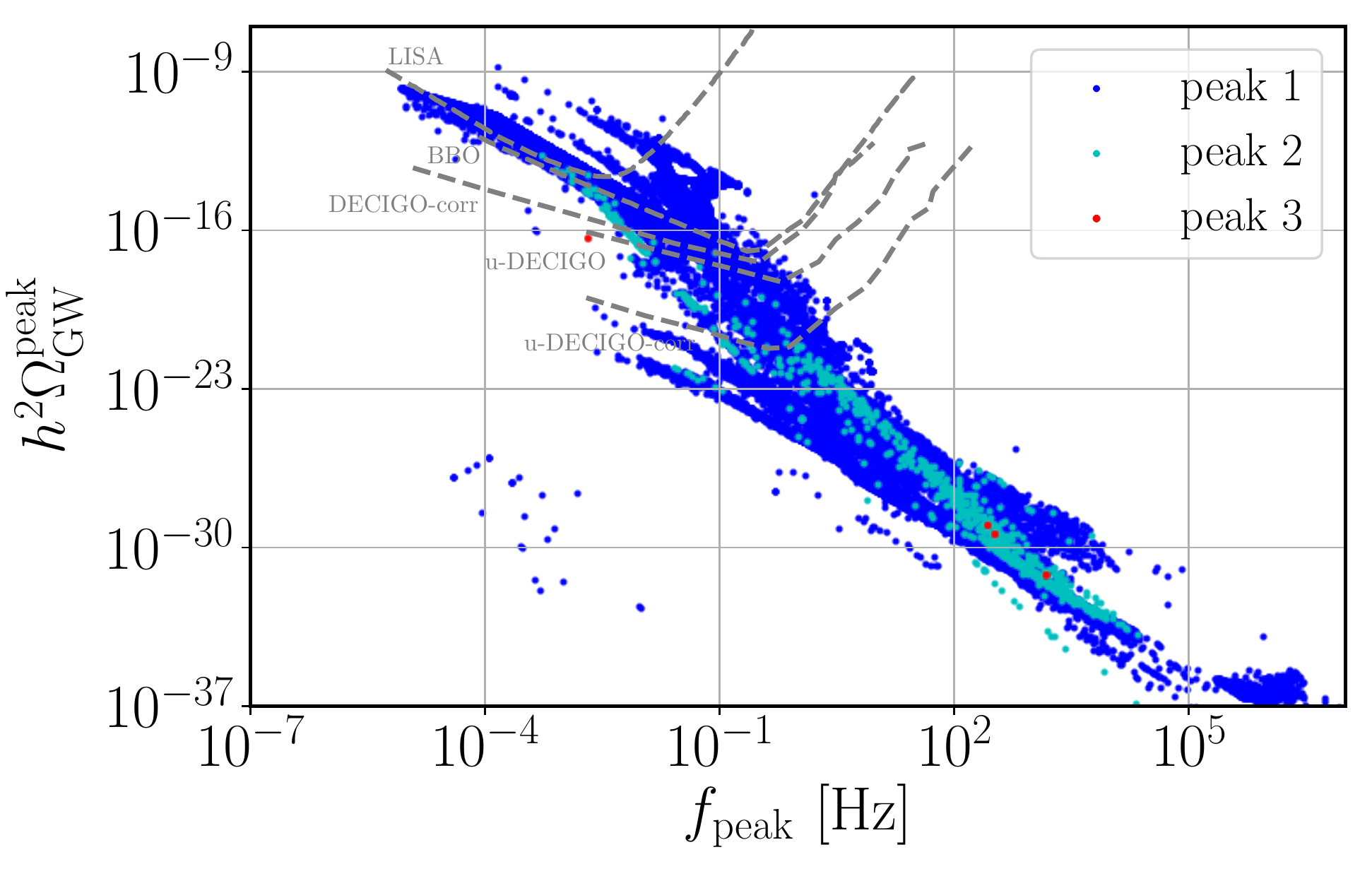}
\caption[a]{\footnotesize Scatter plot showing the number of peaks for given model parameter choices. 
Notice the appearance of double- and even triple-peak features.}
         \label{fig:multi-peak}
\end{subfigure} 
 \end{tabular}
 \end{center}
 \caption{ \footnotesize  The multi-peak feature arising from different phase transitions 
 in the cosmological history of the Universe is very generic in the inverse seesaw with majoron. }
\label{fig:maj-inv-seesaw}
\end{figure}

Multi-peak configurations in the GW spectrum occur very frequently in the inverse seesaw with majoron. This fact is further 
highlighted in Fig.~\ref{fig:maj-inv-seesaw}. Indeed, the double-peak feature of the GW spectrum is a generic prediction of 
our model, that can arise for many parameter choices, as shown in Fig.~\ref{fig:d-peak-line}. Configurations with larger 
peak multiplicities are also possible, as seen in Fig.~\ref{fig:multi-peak}, where the color denotes the peak number, 
$1$ (blue), $2$ (cyan) or $3$ (red). Such a rich structure of the GW spectrum is favoured for relatively large quartic 
couplings involving $\sigma$. From Fig.~\ref{fig:multi-peak} we also see that the GW spectra with three peaks are rarer 
than single or double GW-peak spectra. We also note that a significant fraction of the single peak cases are potentially 
testable at LISA and BBO. However, finding a well-resolved double- or triple-peak feature where more than a single peak could 
be observable in a measurement appears to be very challenging. No single point with such a feature has been 
found by our scans.
As seen in Fig.~\ref{fig:d-peak-line} in most cases both peak-amplitudes occur below the projected sensitivities 
of any planned measurements rendering the observability of the corresponding scenarios very remote. In a subset 
of cases, only the largest peak has been found in the sensitivity ranges of the planned BBO and u-DECIGO missions, 
while the second peak
typically lies outside of the reach of future detectors. This is a direct consequence of the not-so-large quartic 
couplings that are restricted to be smaller than five. While larger values of the quartic couplings could generate
well-separate and measurable double-peak signatures, these scenarios may suffer from larger underlined theoretical
uncertainties so we decided not to discuss them here.
\begin{figure}[htb]
\centering
\begin{center}
\includegraphics[height=6cm,width=8cm]{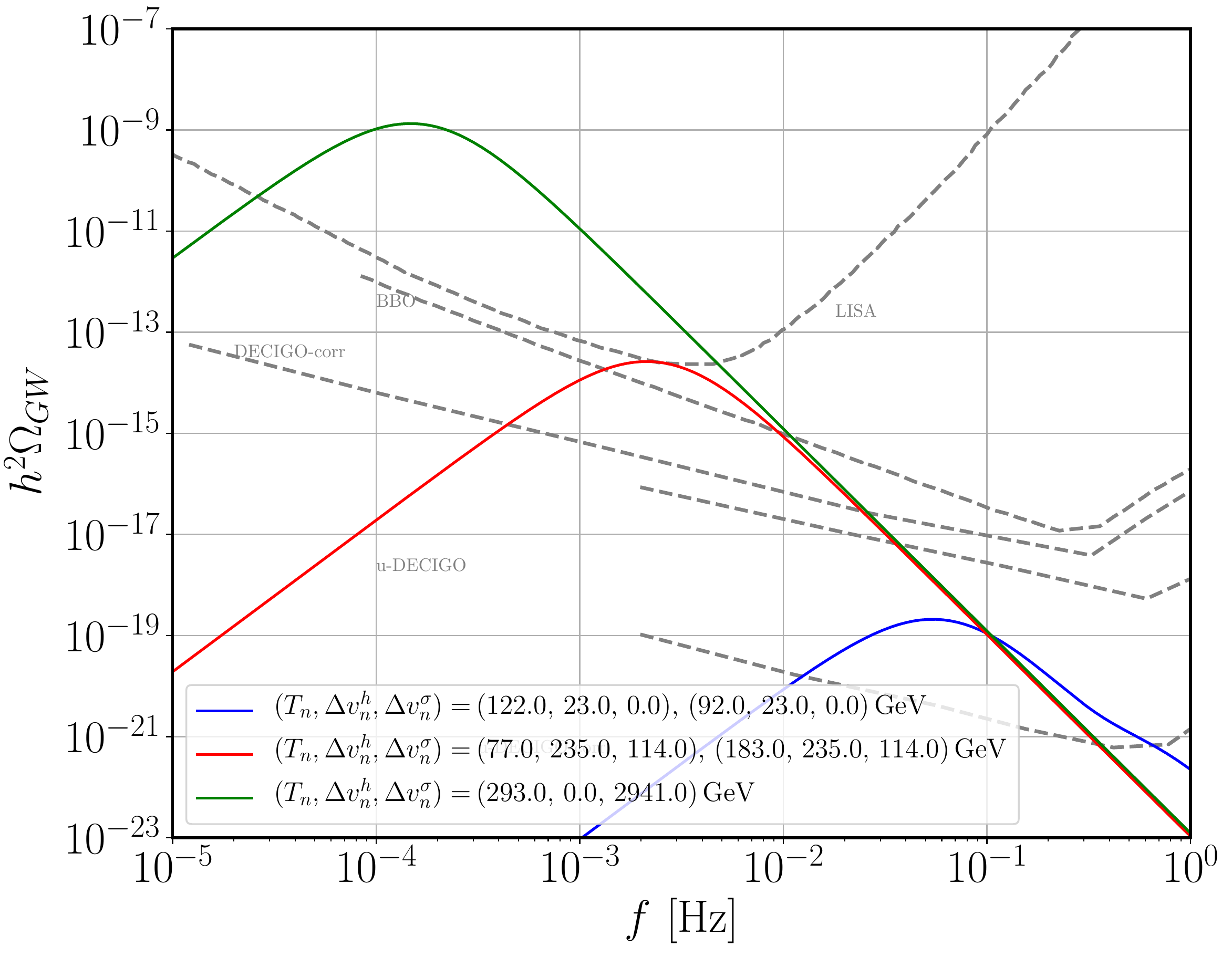}
 \end{center}
\caption{\footnotesize Inverse-seesaw-with-majoron benchmark GW spectra in the scenario with soft $U(1)_L$ 
symmetry breaking (i.e.~with vanishingly small $v_\sigma(T=0) \to 0$). The green curve represents the case with 
single-step FOPT, while the blue and red curves correspond to double-step FOPTs whose characteristics 
are given in Tables~\ref{tab:Bench1-1} and \ref{tab:Bench1-2}. For double-step transitions, the $(T_n,v_n)$ pairs 
in each peak are ordered from low to high frequencies. Here, $\Delta v_n^h = |v_h^f - v_h^i|$ and
$\Delta v_n^\sigma = |v_\sigma^f - v_\sigma^i|$ at a given nucleation temperature $T=T_n$.}
\label{fig:d-peak}
\end{figure}
\begin{table}[htb]   
	\begin{center}
		\begin{tabular}{|c|c|c|c|c|c|}
			\hline                     
			Peak Id & $T_n$ & $\left(v_h^i,v_\sigma^i\right) \to \left(v_h^f,v_\sigma^f\right)$ & $\alpha$ & $\beta/H$	\\    
			\hline
			Green 1 & 293 & $(0,0)   \to (0,2941)$   & $ 0.5$             &  $4.9$ 	\\
			\hline
			Red 1  	& 183 & $(0,114) \to (235,0)$	 & $7.7\times 10^{-4}$ & $7.2 \times 10^4$	\\
			Red 2 	& 77  & $(0,114) \to (235,0)$	 & $0.1$               & $231$              \\
			\hline
			Blue 1 	& 122 & $(193,0) \to (216,0)$	 & $1.4\times 10^{-2}$ & $3.1 \times 10^3$ \\
			Blue 2  & 92  & $(193,0) \to (216,0)$	 & $9.4\times 10^{-3}$ & $3.0 \times 10^4$ \\
			\hline
		\end{tabular} 
\caption{\footnotesize Phase transition parameters for the three curves in Fig.~\ref{fig:d-peak}. 
In ``peak Id'' column, the numbering of multi-step scenarios is ordered from high to low nucleation 
temperature $T_n$, given in units of GeV. The vevs before the transition ($v_{h,\sigma}^{i}$)
and after the transition ($v_{h,\sigma}^{f}$) are given in units of GeV.}
\label{tab:Bench1-1}  
\end{center}
\end{table} 
\begin{table}[htb]     
	\begin{center}
		\begin{tabular}{|c|c|c|c|c|c|c|}
			\hline                     
		Curve & $m_{h_2}$ & $m_A$ & $\lambda_{\sigma h}$  &  $\lambda_\sigma$  & $M$ &  $Y_{\sigma}$	\\    
		\hline
		Green    & $236$		        & $708$                   & $1.7$	          &  $5 \times 10^{-3}$ &  $380$   &  $2$	\\
		\hline
		Red  	 & $192$	        	& $970$                   & $2.3$	          &  $1.5$              &  $93$    &  $0.1$ \\
		\hline
		Blue  	 & $325$	        	& $169$                   & $4$               &  $2.7$              &  $158$   &  $0.1$\\
		\hline
		\end{tabular} 
\caption{\footnotesize Model parameters for the three curves in Fig.~\ref{fig:d-peak}. 
Mass parameters are given in units of GeV.}
\label{tab:Bench1-2}  
	\end{center}
\end{table} 

In order to understand the characteristic features of the GW spectra in the majoron inverse seesaw, 
in Fig.~\ref{fig:d-peak} we show the GW energy density spectrum obtained for distinct nucleation temperatures $T_n$.
Here, we have depicted three benchmark scenarios (for the case of vanishing $v_\sigma$) originating from the single-step 
FOPT (green line) as well as double-step FOPTs i.e.~with two consecutive strong phase transitions (red and blue lines). 
The corresponding values for the model parameters are given in Tab.~\ref{tab:Bench1-1} and Tab.~\ref{tab:Bench1-2}. 

The green curve represents a single-peak scenario with a single very strong $U(1)_L$ phase transition, 
$\Delta v_\sigma(T_n)/T_n \simeq 10$, and softly-broken lepton symmetry. It features a very strong FOPT 
but not EW one. In fact, this peak is a probe of $U(1)_L$ breaking at finite temperatures 
while the associated EWPT is very weak or even of second order. This is possible due to relatively 
large values of a quartic (portal) coupling, $\lambda_{\sigma h}\simeq 2$, and majoron-neutrino Yukawa 
coupling, $Y_\sigma\simeq 2$, which make the $m/T$-ratio sizable. Hence, the cubic $(m/T)^3$ terms in 
the thermal expansion can produce a potential barrier between both vacua, inducing this type of transitions.

On another hand, the other two benchmark points, which have the double-peak feature, have strong EWPT 
for both peaks in the red curve and for the smaller, almost invisible, peak of the blue curve.
Besides this, the blue curve has no $U(1)_L$ breaking in any of the minima of the effective potential at the corresponding nucleation temperatures whereas the red curve exhibits 
a strong $U(1)_L$ breaking for the higher peak and a weak $U(1)_L$ breaking for the hidden peak.
Similarly to the single-peak scenario, the observable peaks in the red and blue curves are generated 
due to large $\lambda_{\sigma h}$ and $\lambda_\sigma$ couplings. Note, the single-peak case (green line) 
lies well within the LISA range \cite{Caprini:2019egz,Caprini:2015zlo,Audley:2017drz}, 
while only the largest peaks in the double-step FOPT scenarios are within the planned sensitivity range of 
the BBO (red line) and u-DECIGO-corr (blue line) measurements \cite{Kudoh:2005as}, respectively.
\begin{figure}[htb]
\begin{center}
 \begin{tabular}{lr}
\begin{subfigure}[b]{0.45\textwidth}
\includegraphics[height=5.5cm,width=8.0cm]{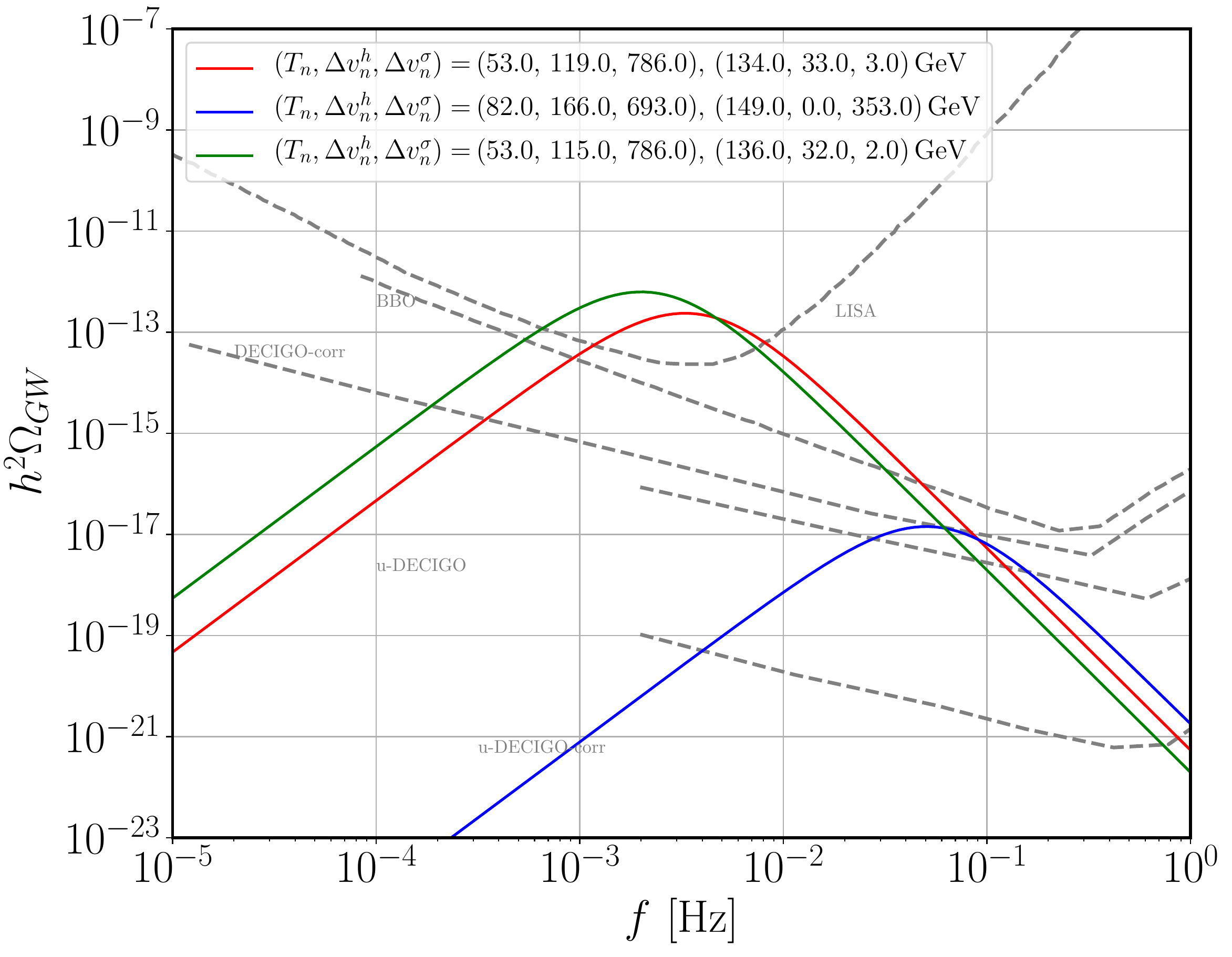}
\caption[a]{\footnotesize The expected GW spectra.
}
         \label{fig:GW-typeI}
\end{subfigure}
\\
\begin{subfigure}[b]{0.45\textwidth}
\includegraphics[height=5cm,width=8.0cm]{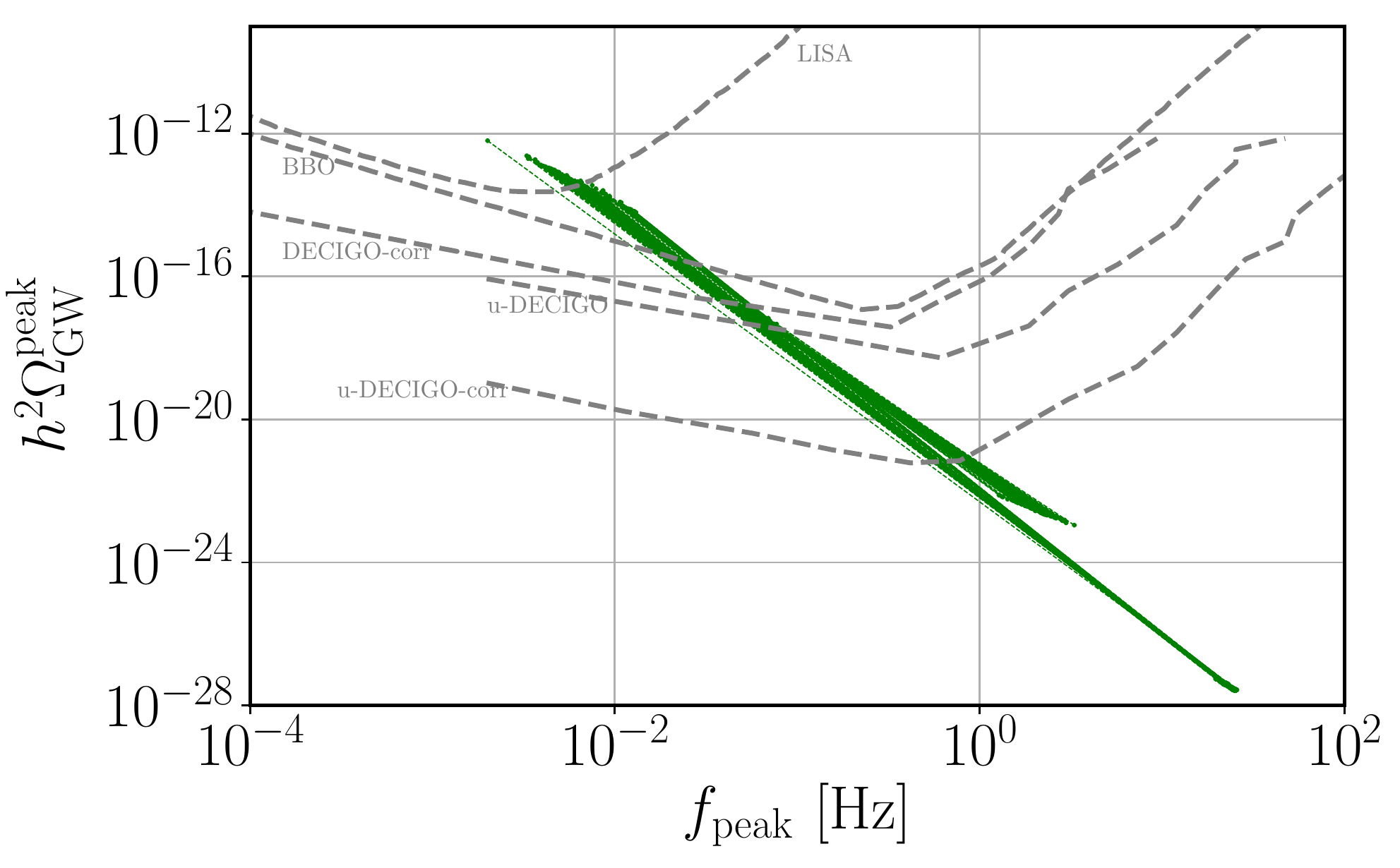}
\caption[a]{\footnotesize Scatter plot showing typical double-peak scenarios.}
         \label{fig:multi-peak-2}
\end{subfigure} 
 \end{tabular}
 \end{center}
\caption{ \footnotesize Gravitational footprints of ``fake'' low-scale seesaw with majoron. 
In both plots we take $v_\sigma \sim  \mathcal{O}(100)$ GeV -- $\mathcal{O}(1)$ TeV.
}
\label{fig:maj-typeI}
\end{figure}
\begin{table}[htb] 
	\begin{center}
		\begin{tabular}{|c|c|c|c|c|}
			\hline                     
			Peak Id    & $T_n$     & $\left(v_h^i,v_\sigma^i\right) \to \left(v_h^f,v_\sigma^f\right)$  &  $\alpha$ & $\beta/H$  \\    
			\hline
			Green 1    & $136$     & $(0,921) \to (32,919)$	   & $9.4 \times 10^{-5}$   &  $1.2 \times 10^6$   	\\
			Green 2    & $53$	   & $(245,786) \to (360,0)$ & $0.5$                  &  $378$    	        \\
			\hline
			Red 1  	   & $134$     & $(0,922) \to (33,919)$  & $10^{-4}$              &  $1.1 \times 10^6$   	\\
			Red 2 	   & $53$      & $(245,786) \to (364,0)$   & $0.5$                  &  $612$              \\
			\hline
			Blue 1     & $149$     & $(0,353) \to (0,0)$	   & $1.5\times 10^{-3}$    & $1.0 \times 10^5$     \\
			Blue 2 	   & $82$      & $(205,693) \to (40,0)$ & $0.03$                 & $4.9 \times 10^3$     \\
			\hline
		\end{tabular} 
		\caption{\footnotesize FOPT parameters for the three curves in Fig.~\ref{fig:GW-typeI}.
			The vevs before ($v_{h,\sigma}^{i}$) and after ($v_{h,\sigma}^{f}$) 
			the phase transition are given in units of GeV.}
		\label{tab:Bench2-1}  
	\end{center}
\end{table} 
\begin{table}[htb]  
	\begin{center}
		\begin{tabular}{|c|c|c|c|c|c|c|c|c|c|}
			\hline                     
			Curve & $m_{h_2}$ & $m_A$  & $\lambda_h $ & $\lambda_{\sigma h}$  &  $\lambda_\sigma$ & $\theta$ & $v_\sigma$ & $Y_{\sigma,1}$ & $Y_{\sigma,2}$	\\
			\hline
			Green     	& $203$ & $188$ & $0.14$ &  $0.03$ &  $0.03$ &  $0.26$ & $790$ & $0.07$ & $1.58$	\\
			\hline
			Red  		& $206$	& $188$ & $0.14$ &  $0.03$ &  $0.03$ &  $0.26$ & $790$ & $0.08$ & $1.59$   \\
			\hline
			Blue  		& $205$	& $188$ & $0.14$ &  $-0.02$ & $0.03$ & $-0.18$ & $790$ & $0.08$ & $1.59$   \\
			\hline
		\end{tabular} 
		\caption{\footnotesize Model parameters for the three curves in Fig.~\ref{fig:GW-typeI}. 
		Masses and $v_\sigma$ are given in units of GeV.}
		\label{tab:Bench2-2}  
	\end{center}
\end{table} 


\subsection{Type-I seesaw with majoron: large $v_\sigma$}
\label{Sec:vs_large}

To further illustrate the importance of future GW measurements for probing neutrino mass generation mechanisms let us consider a type-I seesaw
with majoron where the heavy neutrino mass scale $M$ is generated via a large majoron vev $v_\sigma$ spontaneously breaking the lepton number symmetry. 

As mentioned above, it is clear that, if $Y_\nu \sim \mathcal{O}(1)$, then $M = Y_\sigma v_\sigma/\sqrt{2} \sim \mathcal{O}(10^{14})$ GeV, hence 
$v_\sigma \sim \mathcal{O}(10^{14})$ GeV for $Y_\sigma \sim \mathcal{O}(1)$. In this limit, all the new particles 
can be integrated out for processes occurring at the EW scale\footnote{All the physical majoron couplings 
are highly suppressed yielding no effect on EW scale physics.}, leading to no-FOPT solutions.

However, one can take $Y_\nu \sim \mathcal{O}(10^{-6})$, corresponding to $M = Y_\sigma v_\sigma/\sqrt{2} 
\sim \mathcal{O}(100)$ GeV. The majoron and neutrino fields do not decouple at the EW scale in this case, 
and can still lead to strong FOPTs -- hence to potentially observable primordial GW signals. One sees that 
this ``fake'' low-scale seesaw scenario requires tiny values of the neutrino ``Dirac'' Yukawa couplings $Y_\nu$ to fit the small 
neutrino masses in the presence of relatively large $U(1)_L$ symmetry breaking scale $v_\sigma$ that can 
be placed not too far from the EW scale. Such ``fake'' low-scale seesaw contrasts with the ``genuine'' 
low-scale seesaw considered in Sec.~\ref{Sec:vs_0}, which does not require this restriction. 

Here, for an easy one-to-one comparison with the ``genuine'' inverse seesaw scenario studied above, we will consider a type-I 
majoron seesaw model with six heavy neutrinos. This way we preserve 
the number of fermionic degrees of freedom entering the thermal corrections by considering the following Lagrangian
\begin{eqnarray}
   \mathcal{L}_{\rm Yuk}^{\rm Type-I} & = & Y_{\nu,i} \bar{L} H \nu_i^c + Y_{\sigma,i} \sigma \nu_i^c \nu_i^c + h.c.
 \label{lag-typeI}
\end{eqnarray}
where $\nu^c$ and $S$ in the inverse seesaw scenario are replaced by $\nu_{1}^c$ and $\nu_2^c$ in this extended type-I 
seesaw variant, and the off-diagonal terms in the heavy neutrino mass matrix and ``Dirac'' Yukawa couplings $Y_{\nu,i}$ 
are assumed to be negligibly small.
As our main result, we find that, in a large region of parameter space, both ``fake'' and ``genuine'' low-scale 
seesaw $+$ majoron lead to the possibility of strong FOPTs. The corresponding GW spectra in the ``fake'' seesaw 
obtained for $v_\sigma \sim \mathcal{O}(100)$ GeV  -- $\mathcal{O}(1)$ TeV are shown in Fig.~\ref{fig:maj-typeI}. 
Parameter values of this model associated to Fig.~\ref{fig:GW-typeI} are given in Tab.~\ref{tab:Bench2-2} 
and Tab.~\ref{tab:Bench2-1}.

\section{Conclusion}
\label{Sec:Conclusion}

To conclude, we analysed the most popular implementations of the type-I seesaw mechanism for neutrino mass 
generation. We studied both the cases of explicit and spontaneous breakdown of the lepton number symmetry 
in the neutrino sector. The second, ``dynamical'' symmetry breaking implies the majoron field. We have 
found that various scenarios lead to different patterns of phase transitions. We showed that explicit 
lepton number violation in the neutrino sector cannot induce any strong electroweak phase transition. 
Therefore, it does not lead to any gravitational-wave background signal testable by next-generation 
satellite interferometers.

The case when neutrino masses emerge from a dynamical mechanism in which lepton number violation happens 
spontaneously leads to much clearer gravitational footprints. Within such majoron seesaw case, we found 
that both the standard type-I seesaw (taken at a low scale) and the ``genuine'' low-scale type-I seesaw 
(like the inverse seesaw) predict a strong gravitational wave signal, testable in the $0.1 - 100\, {\rm mHz}$ 
frequency range. This highly motivates future experimental proposals, including LISA, u-DECIGO and BBO missions, 
accessing to the mHz frontier, as an indirect and complementary probe of neutrino mass generation, providing 
an important information on the electroweak phase transition. 

While ``genuine'' low-scale seesaw would also predict large charged \lfv \cite{Bernabeu:1987gr,Branco:1989bn, Rius:1989gk,Deppisch:2004fa,Antusch:2006vw}, and unitarity violation in neutrino 
propagation~\cite{Escrihuela:2015wra, Forero:2011pc,Miranda:2016wdr}, these features are absent 
in the ``fake'' low-scale seesaw. This way one can distinguish the two schemes in high-intensity/energy 
frontier setups. Here we have shown that ``fake'' and ``genuine'' schemes may also have potentially distinct 
gravitational footprints. We saw explicitly that they can produce different gravitational-wave spectra, testable 
in upcoming gravitational-wave experiments. As we stand right now, the new portal provided by 
the gravitational-wave physics in the multi-messenger era may contribute to shed light on the mystery 
of neutrino mass generation.

\acknowledgements 
\noindent

Useful discussions and correspondence with Zurab Berezhiani, Yifu Cai
and Nico Yunes are gratefully acknowledged.
A.P.M wants to thank Marek Lewicki and Bogumi\l a \'Swie$\dot{\rm{z}}$ewska
for insightful discussions about bubble wall collision contributions to the
spectrum of GW. A.A. and A.M. wish to
acknowledge support by the NSFC, through grant No. 11875113, the
Shanghai Municipality, through grant No. KBH1512299, and by Fudan
University, through grant No. JJH1512105. J.B. acknowledges his
partial support by the NSFC, through the grants No. 11375153 and
11675145.  A.A. and A.M. would like to thank IFIC for hospitality
during the preparation of this work.  R.P.~is supported in part by the
Swedish Research Council grants, contract numbers 621-2013-4287 and
2016-05996, as well as by the European
Research Council (ERC) under the European Union's Horizon 2020
research and innovation programme (grant agreement No 668679).  The
work of R.P. was also supported in part by the Ministry of Education,
Youth and Sports of the Czech Republic, project LT17018. The work of
A.P.M. has been performed in the framework of COST Action CA16201 
``Unraveling new physics at the LHC through the precision frontier'' 
(PARTICLEFACE). A.P.M.~is supported by Funda\c{c}\~ao para a 
Ci\^encia e a Tecnologia (FCT), within project UID/MAT/04106/2019 (CIDMA) 
and by national funds (OE), through FCT, I.P., in the scope
of the framework contract foreseen in the numbers 4, 5 and 6 of 
the article 23, of the Decree-Law 57/2016, of August 29, changed 
by Law 57/2017, of July 19.~A.P.M.~is also supported by the 
\textit{Enabling Green E-science for the Square Kilometer 
Array Research Infrastructure} (ENGAGESKA), POCI-01-0145-FEDER-022217, 
and by the project \textit{From Higgs Phenomenology to the Unification 
of Fundamental Interactions}, PTDC/FIS-PAR/31000/2017.
R.V. and J.W.F.V. are supported by the Spanish grants SEV-2014-0398
and FPA2017-85216-P (AEI/FEDER, UE), PROMETEO/2018/165 (Generalitat
Valenciana) and the Spanish Red Consolider MultiDark
FPA2017-90566-REDC.


\bibliographystyle{utphys}
\bibliography{bibliography}

\end{document}